\PassOptionsToPackage{table,dvipsnames}{xcolor}
\documentclass[acmtog,nonacm]{acmart}
\AtBeginDocument{%
  \providecommand\BibTeX{{%
    \normalfont B\kern-0.5em{\scshape i\kern-0.25em b}\kern-0.8em\TeX}}}

\setcopyright{acmcopyright}
\copyrightyear{2018}
\acmYear{2018}
\acmDOI{XXXXXXX.XXXXXXX}

\acmConference[Conference acronym 'XX]{Make sure to enter the correct
  conference title from your rights confirmation emai}{June 03--05,
  2018}{Woodstock, NY}
\acmBooktitle{Woodstock '18: ACM Symposium on Neural Gaze Detection,
 June 03--05, 2018, Woodstock, NY} 
\acmPrice{15.00}
\acmISBN{978-1-4503-XXXX-X/18/06}

\usepackage{tikz,graphics,color,caption,subcaption}
\usepackage{hyperref}
\usepackage[linesnumbered,ruled,vlined]{algorithm2e}
\newcommand{\norm}[1]{\left\lVert#1\right\rVert}

\begin{document}
\graphicspath{{graphPath/}}

\title{Automatic Skinning using the Mixed Finite Element Method}


\author{Hongcheng Song}
\email{songhong@usc.edu}
\orcid{1234-5678-9012}
\affiliation{%
  \institution{University of Southern California}
  \streetaddress{P.O. Box 1212}
  \city{Los Angeles}
  \state{California}
  \country{USA}
  \postcode{43017-6221}
}

\author{Dimitry Kachkovski}
\affiliation{%
  \institution{Unity Technologies/The Open University}
   \city{Vancouver}
    \country{Canada}}
  \orcid{1234-5678-9012}
\email{dimitry.kachkovski@gmail.com}

\author{Shaimaa Monem}
\email{monem@mpi-magdeburg.mpg.de}
\affiliation{%
  \institution{Max Planck Institute for Dynamics of Complex Technical Systems, Institute of Mathematical
Optimization OVGU}
  \city{Magdeburg}
  \country{Germany}
  \postcode{39106}
}

\author{Abraham Kassahun Negash}
\affiliation{%
  \institution{Addis Ababa Institute of Technology}
  \city{Addis Ababa}
  \country{Ethiopia}}
\email{abknegash@gmail.com}

\author{David I.W. Levin}
\orcid{1234-5678-9012}
\affiliation{%
  \institution{University of Toronto}
  \streetaddress{1 Th{\o}rv{\"a}ld Circle}
  \city{Toronto}
  \country{Canada}}
\email{diwlevin@gmail.com}

\renewcommand{\shortauthors}{Trovato and Tobin, et al.}

\begin{abstract}

In this work, we show that exploiting additional variables in a mixed finite element formulation of deformation leads to an efficient physics-based character skinning algorithm. Taking as input, a user-defined rig, we show how to efficiently compute deformations of the character mesh which respect artist-supplied handle positions and orientations – but without requiring complicated constraints on the physics solver, which can cause poor performance. Rather we demonstrate an efficient, user controllable skinning pipeline that can generate compelling character deformations, using a variety of physics material models.

\end{abstract}



\begin{teaserfigure}
  \includegraphics[width=\textwidth]{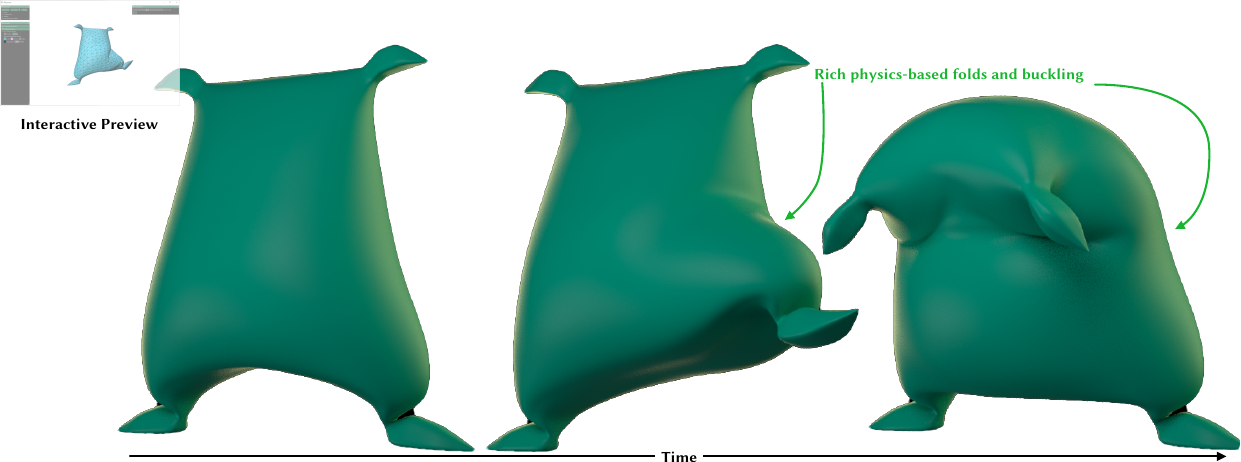} 
  \caption{Our mixed finite element skinning technique replaces precomputed skinning weights with efficient simulation \textbf{(1.7 sec/frame using numpy)}, yielding character deformations featuring appealing physics-based detail, such as volume conservation, folding and buckling.  \label{fig:teaser}}
\end{teaserfigure}

\received{20 February 2007}
\received[revised]{12 March 2009}
\received[accepted]{5 June 2009}

\maketitle

\section{Introduction}

Character skinning, the specification of character mesh motion via the manipulation of a sparse set of artist-controlled handles, is the defacto standard methodology for character animation in computer graphics. In a typical workflow an artist manually specifies the positions of a set of direct manipulation handles (known as the rig) along with a set of precomputed per-vertex weights. These weights propagate rig motion to the surface vertices of the character yielding the final, deformed mesh pose. Commensurate with the importance of skinning in the graphics workflow, many highly successfully algorithms have been proposed for both automatic generation of rigs and of skinning weights.

The most common implementation of skinning, called linear blend skinning (LBS), computes mesh deformation as a linear combination of scalar weight functions and rigid or affine handle motions. This method is fast and, skilled rig artists can produce compelling and realistic animations, often with significant time spent crafting weight functions. To alleviate this a number of methods for computing weights automatically have been proposed. These methods attempt to generate weights that produce physically plausible character deformations. More recent techniques, inspired by the observation that skinning algorithms are essentially trying to approximate the results of a physics-based animation, directly incorporate such algorithms into the animation pipeline (at the cost of increased computational complexity). 

Inspired by the use of physics-based animation for skinning, this paper attempts to do away with weight generation for skinning entirely -- replacing it with a physics simulation, but without introducing unmanageable computational burden. Standard approaches to elasticity simulation for deformation rely on variants of the finite element method (FEM) which solve a large (non-linear, as soon as rotations are involved) system of equations to compute deformed mesh positions. Applications of  FEM  to the skinning problem are difficult as specifying the necessary constraints to force the simulation to track the rig motion can be overly constraining and produce undesirable artifacts. 

Instead, we exploit a mixed finite element method (MFEM) discretization which introduces extra rotation and strain variables per mesh element. By explicitly tying these additional variables to rig handles we can avoid the over-constraining issues incumbent in a standard FEM approach. Further, because the MFEM discretization allows us to directly fix element rotations based on the rig, useful deformation models such as As-Rigid-As-Possible (ARAP) or Co-Rotated Linear Elasticity become linear (without inducing the standard inflation artifacts) requiring only a single linear solve per frame. This can be accomplished interactively in our python implementation of the algorithm. Along with a reduction in artifacts and an increase in performance, our MFEM skinning approach requires no user specified skinning weights and retains the benefits of having a physics simulator in the animation loop including the ability for artists to specify different material properties for a mesh, different material models, and to respond to external perturbations like contact.  

\section{Related work}

Linear blend skinning (LBS) is the most common form of skinning applied to character animation. LBS expresses deformation of mesh vertices as a linear combination of rigid or affine handle motions and a set of scalar weighting functions~\citep{skinningcourse2014}. Rigs and weights are often artist-designed however automated methods exist for both tasks~\citep{RigNet,Baran_Skinning_2007}.

In this work we focus on the problem (rather the avoidance of) skinning weight generation and so focus our review on similarly motivated methods. A popular set of approaches to automatic skinning weight generation are partial differential equation-based (PDE) methods. These approaches use a variational approach to design skinning weights that inherit smoothness properties from an underlying PDE~\citep{Baran_Skinning_2007,BBW:2011,Wang_subspace_2015}. Such weights can produce plausible deformations but their smooth, scalar nature means they cannot replicate important physical deformations such as volume preservation and buckling. Since they are precomputed, they cannot respond to external perturbations between the character and the environment.

Alternative variational approaches directly solve an optimization problem to compute deformations~\citep{Shaefer_MLS_2006}. Physics-based methods for computing character deformation can be viewed through this lens since they solve constrained optimization problems using physical energies. \citet{McAdams_Skinning_2011} has the most influence on our work, as they also replace skinning weight calculation with physics simulation. However, our particular formulation is significantly more flexible (doesn't require a hexahedral mesh or specialized quadrature scheme) , efficient (co-rotated elasticity models require only a single linear solve per-timestep) and easier to implement (no extreme optimization required for interactive performance) than their impressive, but highly specialized approach. Also closely related, \cite{Jacobson12} combines a reduced space discretization with a local-global solver to produce fast, automatic skinning transformations. Different from us, their local-global approach is limited to the ARAP~\cite{Sorkine07} material model and their use of a reduced space precludes computing deformations that respond to local perturbations. Alternate approaches layer dynamics on top of already animated meshes, maintaining skinning-physics orthogonality to maintain artist-intent~\citep{Zhang_CompDynamics_2020}. Such approaches solve a different problem than us, focusing on augmenting an existing animation rather than using physics methods to synthesize entirely new deformations. 

Our method is based on the mixed finite element method (MFEM) for elasticity~\citep{Trusty22} which directly exposes rotations and deformations as mixed variables. We depart from other similar formulations~\citep{brown2021wrapd} for mesh deformation by exploiting the artist-provided rotations of each rig handle and properly coupling to a rig rather than just point constraints. By taking these variables as knowns in the MFEM formulation we reduce the nonlinear deformation problem from  non-linear to linear for  common material models. Thus our method only requires a single linear solve per frame compared to the alternating projection, ADMM, or newton's methods applied in prior work while still inheriting the favourable early convergence properties of an MFEM discretization~\citep{Trusty22}. Taken together our work provides a simple, efficient method for generating physically plausible skinning deformations without  requiring manual skinning weight specification.

\section{Method}

Following the standard approach, we model object deformation via its strain energy density function $\boldsymbol{\Psi}$ \cite{Eftychios12}. Total object potential energy $\mathnormal{E}$ is given by integrating $\boldsymbol{\Psi}$ over the entire undeformed object domain $\Omega$
\begin{equation}
    \mathnormal{E}[\phi(X)] = \int_{\Omega} \Psi(\boldsymbol{F}(\boldsymbol{X}))\, \mathrm{d}\Omega
    \label{eq:potential}
\end{equation}

where $\phi(X)$ is the deformation map function that maps an arbitrary point $\boldsymbol{X}$ in undeformed space to $\boldsymbol{x}$ in world space, $\boldsymbol{\mathnormal{F}} = \nabla\phi(\boldsymbol{X}) \in \mathbb{R}^{3\times3}$ is the deformation gradient and $\Omega$ is the 3D undeformed domain. 


Character skinning can be considered an elatostatics problem wherein the goal is to minimize equation~\ref{eq:potential}. However, in standard discretizations only vertex variables are exposed for user control -- constraining these variables to follow affine skinning input can lead to artifacts. A mixed finite element formulation which exposes additional rotation and deformation variables mitigates this issue.

The mixed form of Equation~\ref{eq:potential} is written as 
\begin{equation}
    \mathbb{E} = \int_{\Omega} \boldsymbol{\Psi}(\boldsymbol{\mathnormal{S}}) + \mathnormal{C}(\boldsymbol{\mathnormal{X}})\, \mathrm{d}\Omega
    \label{eq:mfem}
\end{equation}
\begin{equation}
    s.t.\; \nabla\phi(\boldsymbol{X}, \, \boldsymbol{\mathnormal{S}}) = \boldsymbol{\mathnormal{R}}(\boldsymbol{\mathnormal{H}}, \, \boldsymbol{\mathnormal{X}})\boldsymbol{\mathnormal{S}}
\end{equation}

where $\mathnormal{C}(\boldsymbol{\mathnormal{X}})$ are some additional applied constraints, $\boldsymbol{\mathnormal{S}} \in \mathbb{R}^{3\times3}$ is a symmetric deformation matrix, which now becomes our second independent variable, and $\boldsymbol{\mathnormal{R}}(\boldsymbol{\mathnormal{H}}, \, \boldsymbol{\mathnormal{X}})$ is the rotation matrix  applied exclusively using the set $H = \{h_1, h_2, ..., h_n\}$ containing all handles/joints on the skeleton. We can then reformulate equation~\ref{eq:mfem} via Lagrange multipliers as
\begin{equation}
    \mathbb{E} = \int_{\Omega} \boldsymbol{\Psi}(\boldsymbol{\mathnormal{S}}) + \mathnormal{C}(\boldsymbol{\mathnormal{X}}) + \mathnormal{T}(\boldsymbol{\mathnormal{X}}, \,\boldsymbol{\mathnormal{S}})\, \mathrm{d}\Omega
    \label{eq:lm}
\end{equation}

where $\boldsymbol{\mathnormal{\Lambda}}(\boldsymbol{\mathnormal{X}}) \in \mathbb{R}^{3 \times 3}$ is the Lagrangian multiplier matrix and $\mathnormal{T}(\boldsymbol{\mathnormal{X}}, \,\boldsymbol{\mathnormal{S}}) = \boldsymbol{\mathnormal{\Lambda}}(\boldsymbol{\mathnormal{X}}) : (\nabla\phi(\boldsymbol{X}) - \boldsymbol{\mathnormal{R}}(\boldsymbol{\mathnormal{H}})\boldsymbol{\mathnormal{S}})$. Although introducing new variables increases the total degrees of freedom, this mixed form formulation gives us access to explicitly manipulate elements' rotation through skeleton handles; we then solve a standard minimization problem via equation ~\ref{eq:lm} to compute the full deformation. In our discrete setting, we define $\boldsymbol{\mathnormal{F}}(\boldsymbol{\mathnormal{X}}) = \boldsymbol{\mathnormal{B}} \boldsymbol{\mathnormal{x}}$ where $\boldsymbol{\mathnormal{B}} \in \mathbb{R}^{3 \times 4}$ is the deformation gradient operator.

For each element $k$, we have a symmetric deformation matrix, a rotation matrix and a Lagrangian multiplier as $\boldsymbol{\mathnormal{S}}_{\mathnormal{k}}$, $\boldsymbol{\mathnormal{R}}_{\mathnormal{k}}$ and $\boldsymbol{\mathnormal{\Lambda}}_{\mathnormal{k}}$ respectively. Therefore, our full mixed energy discretization $U$ takes the form 
\begin{equation}
    \mathnormal{U} = \sum_k (\boldsymbol{\Psi}(\boldsymbol{\mathnormal{S}}_{\mathnormal{k}}) + \boldsymbol{\Lambda}_{\mathnormal{k}} : (\boldsymbol{\mathnormal{B}}_{k}\boldsymbol{\mathnormal{x}} - \boldsymbol{\mathnormal{R}}_{k} \boldsymbol{\mathnormal{S}}_{k}))\mathnormal{W}_k + \mathnormal{C}(\boldsymbol{\mathnormal{x}}) 
\end{equation}

where $\mathnormal{W}_k$ is the matrix containing the volumes of the tetrahedrons. Because $\boldsymbol{\mathnormal{S}}_k$ is a symmetric $3\times3$ matrix, we use a $6$-Vector $\boldsymbol{\mathnormal{s}}_k$ to represent it, and a $9$-vector $\boldsymbol{\mathnormal{\lambda}}_k$ to represent the unsymmetric $\boldsymbol{\mathnormal{\Lambda}}_k$. $x$ is the stacked vector of all mesh vertices. Then, our discrete mixed energy can be written in the vectorized form as
\begin{equation}
    \mathnormal{L}(x, s, \lambda) = \sum_k (\boldsymbol{\Psi}(s_k) + \boldsymbol{\lambda}_k (\boldsymbol{Bx} - \boldsymbol{R}_k \boldsymbol{s}_k))w_k + C(\boldsymbol{x})
\end{equation}

where $\boldsymbol{R}_k \in \mathbb{R}^{9 \times 6}$ is the new formulated rotation with respect to a 6-Vector. In our cases, we require a constraint to pin vertices along the skeleton; $C(\boldsymbol{x}) = \frac{ks}{2} \| \boldsymbol{P}\boldsymbol{x} - x_p \|_{2}^2$, where $\boldsymbol{P}$ is a projection matrix, $x_p$ is the pinned vertices and $k_s$ is the constrain parameter. Now, the ultimate goal is to find a root point of our discrete system energy:
\begin{equation}
    x^*, s^*, \lambda^* = \underset{\lambda}{\mathrm{argmax}}\mbox{ }\underset{s, x}{\mathrm{argmin}}\; L(x, s, \lambda)
\end{equation}

where $x \in \mathbb{R}^{3n}$, $s \in \mathbb{R}^{3m}$ and $\lambda \in \mathbb{R}^{3m}$ (n is the number of vertices and m is the number of tetrahedron element).

To solve this equation, we  form the KKT system
\begin{equation}
    \begin{bmatrix}
H_s & 0 & \nabla_{s\lambda}L\\
0 & H_x & \nabla_{x\lambda}L\\
\nabla_{s\lambda}L^T & \nabla_{x\lambda}L^T & 0
\end{bmatrix}
\begin{bmatrix}
    \boldsymbol{s}\\
    \boldsymbol{x}\\
    \boldsymbol{\lambda}
\end{bmatrix} = 
\begin{bmatrix} 
    -g_s\\
    g_x\\
    0
\end{bmatrix}
\end{equation}
where $H_s$ and $H_x$ are the Hessian matrix of $L(\boldsymbol{s},\, \boldsymbol{x}, \boldsymbol{\lambda})$ with respect to $\boldsymbol{s}$ and $\boldsymbol{x}$ respectively, $g_s$ and $g_x$ are $\nabla_sL(x, s, \lambda)$ and $\nabla_xL(x, s, \lambda)$ respectively. In our energy, $\nabla_{s\lambda}L$ and $\nabla_{x\lambda}L$ will always be $\left[-R^T\right]$ and $B^T$; therefore, we can re-write equation [12] as:
\begin{equation}
    \begin{bmatrix}
H_s & 0 & \left[-R\right]^T\\
0 & H_x & B^T\\
\left[-R\right] & B & 0
\end{bmatrix}
\begin{bmatrix}
    \boldsymbol{s}\\
    \boldsymbol{x}\\
    \boldsymbol{\lambda}
\end{bmatrix} = 
\begin{bmatrix} 
    -g_s\\
    g_x\\
    0
\end{bmatrix}
\end{equation}

where $\left[-R^T\right]$ is a block diagonal matrix storing the $9\times6$ matrix $R_k$, and it acts by applying the per element rotation to the symmetric deformation vector to produce a flattened deformation gradient. However, the left hand side matrix may not always be symmetric positive definite, which is hard to solve; therefore, we transform our system to a symmetric positive definite (SPD) form by condensation. We also use the Schur-Complement to reduce our original system to a condensed one, where we eliminate and substitute variables one by one until we only have $x$ left in the system. First, we write
\begin{equation}
    \boldsymbol{s} = H_s^{-1}([R]^T\lambda - g_s).
\end{equation}
We then can then solve for Lagrange multipliers via 
\begin{equation}
    \boldsymbol{\lambda} = ([R]H_s^{-1}[R]^T)^{-1}(B\boldsymbol{x} + [R]H_s^{-1}g_s)
\end{equation}
Finally, we use our $\lambda$ update equation to arrive at the condensed, per-vertex system
\begin{equation}
    (H_x + ([R]^{-1}B)^TH_s[R]^{-1}B) \boldsymbol{x} = g_x -([R]^{-1}B)^Tg_s
    \label{eq:final}
\end{equation}

where we use a fast, batched pseudo inverse to invert the block matrix [R].

Now, we can solve  equation 12 for the new positions of model vertices. Because rotations are specified by the rig, for material energies that are quadratic as a function of deformation, and for which the constraint functions are quadratic, the energy hessian is constant per-frame. Thus for such cases (which encompass useful material models like ARAP and Co-rotated linear elasticity) we require only a single linear solve to reach the global optimum.

\subsection{Final Algorithm}
In a skinning problem, model poses are determined by the orientation of joints or handles. We construct different rotation clusters for different bones by computing distances from the barycenter of a tetrahedron to each bone segment, then assigning each tetrahedron with the closest bone segment; hence, we have piece-wise rotation cluster scheme and all tetrahedrons in the same rotation cluster are assigned with the same rotation. With all tetrahedron being assigned with corresponding rotation, we can build up a new block diagonal rotation matrix $\left[R\right]$ every time the user manipulates the rig handles. We feed this into our condensed system; then, solve for a new shape. 

\begin{algorithm}[]
    
	\textbf{Input:} skeleton $h$, skeleton poses sequence $R$, 3D tetrahedron mesh $X$. \\
    \BlankLine
    $\boldsymbol{x} \leftarrow X$. \\
    \BlankLine
    \tcp*[h]{equation 12} \\
    $\left[H_s,\, g_s\right] \leftarrow$ derivative($\boldsymbol{\Phi}(\boldsymbol{s}),\, \boldsymbol{s}$)\\
    $\left[H_x,\, g_x\right] \leftarrow$ derivative($\boldsymbol{C}(\boldsymbol{x}),\, \boldsymbol{x}$)\\
    $\left[\nabla_sL,\, \nabla_xL\right] \leftarrow$ derivative($\boldsymbol{L}(\boldsymbol{x},\, \boldsymbol{s}),\, \boldsymbol{x},\, \boldsymbol{s}$)\\
	\vspace{0.3cm}
           \For{$R(t)$ in $R$}{
           $\left[R\right] \leftarrow R(t)$\\
           $A \leftarrow$ assemble($H_s,\, H_x,\, \nabla_sL,\, \nabla_xL,\, \left[R\right]$) \tcp*[f]{equation 16} \\
           $b \leftarrow$ assemble($-g_s, g_x, \boldsymbol{0}$)\\
           $x = A^{-1}b$
           \BlankLine
           update and visualize new model \label{eq:final} \\
           }
	\caption{ MFEM algorithm}
	\label{alg: psuedoAlg}
\end{algorithm}

\subsection{Implementation details}

All our experiments are run on a desktop computer with Windows operating system and 8-core Intel i9-11950H 2.60GHz processor with 64GB RAM, using Python 3.10.10. We employed the ftetWild algorithm \cite{ftetWild} for tetrahedralizing all surface meshes in our study. The gradient and Hessian computations relied on Bartels \cite{bartels}. Our code is completely written in Python, leveraging the efficient algorithms provided by numpy \cite{harris2020array} and scipy \cite{2020SciPy-NMeth}. Finally, Polyscope and Maya were used to render our results, and we intend to make our \href{https://github.com/anonymous}{code} available.

\begin{figure}
    \centering
    \includegraphics[width=\columnwidth]{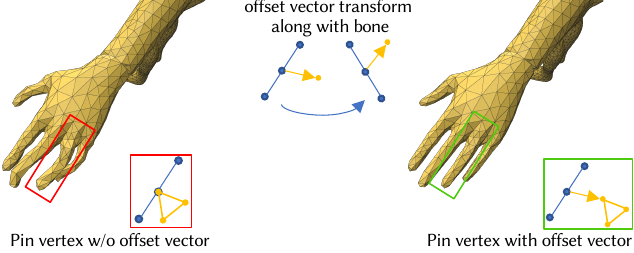}
    \caption{Pinning vertices without offset vector will introduce undesired deformation in neutral shape (left) - and Pinning vertices with offset produce a fine result.}
    \label{fig:offset}
\end{figure}

\paragraph{Initial shape preservation}
In the initial implementation we used the input handle positions as constraints, that is $x_p = H_p$, where $H_p$ are the positions of the handles. However, that led to alterations of the initial positions of the vertices that were chosen to be pinned as they were snapped to the handle positions. Our solution was to constrain the pinned vertices to the handles instead, that is $x_p = H\hat{x}_p$, where $\hat{x}_p$ are the rest pose positions of the pinned vertices. This allowed to preserve the initial state of the provided mesh (Figure~\ref{fig:offset}).

\begin{figure}
    \centering
    \includegraphics[width=\columnwidth]{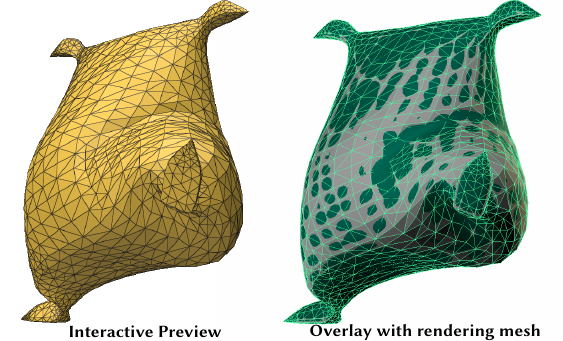}
    \caption{Left: Flour sack pose in our interactive preview application. Right: Flour sack surface mesh overlaid with wrapped rendering surface.}
    \label{fig:workflow}
\end{figure}

\paragraph{Rendering}
We embedded our method into a standard VFX animation pipeline. Surface meshes were exported from our simulation system into Maya where they were automatically wrapped with a rendering mesh for final visualization. Figure~\ref{fig:workflow} shows a side-by-side of a preview mesh along with the mesh, in Maya, overlaid with its surface wrap.

\section{Results}

\begin{figure}
     \centering
     \includegraphics[width=\columnwidth]{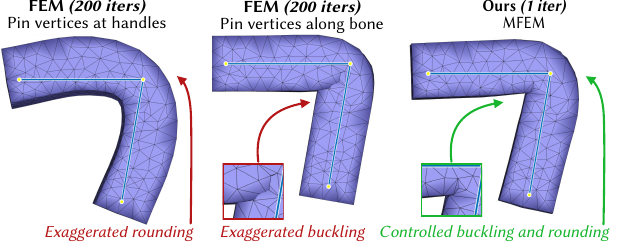}
     \caption{Using vertex constraints to couple standard finite element method to a rig leads to artifacts due to under- (left) or over-constraining (center) the discretization. Our method avoids these issues with the additional advantage of requiring only a single linear system solve.
        \label{fig:90Bend}}
\end{figure}

\paragraph{comparing to standard FEM model} We evaluated our method by subjecting a beam to over 90 degrees of bending and compared it to the solution provided by the standard FEM system, as illustrated in Figure \ref{fig:90Bend} . For the FEM model, we attempted two approaches to pin the vertices around the handles. Initially, we pinned the three closest vertices to each handle, resulting in a pleasing curved shape but not adhering to the defined skeleton configuration. Alternatively, we pinned all vertices along the bones, which preserved the bone structure but led to self-intersections and mesh overlap around the rotated joint. In contrast, our MFEM approach required minimal pinning around the handles and midpoints, resulting in a smoother and rounded bending that accurately preserved the skeleton. Additionally, the execution time of MFEM was significantly faster, highlighting the numerical stability and advantages of employing MFEM over FEM.

\begin{figure}[!h]
     \centering
     \includegraphics[width=\columnwidth]{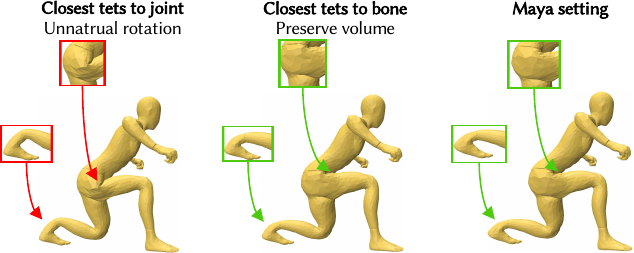}
     \caption{Clustering rotations of tetrahedron by the closest distance to joint produces unnatural deformation at hip and left-leg (left) - using closest distance to each bone segment for rotation clusters gives plausible result (middle) - user can also define rotation clusters on their own (right) which produces similar result as closest bone distance.
        \label{fig:rcluster}}
\end{figure}

\paragraph{Exploring different rotation clustering} Assignment of input rotation to tetrahedron can be approached in multiple ways. In our experiments, we explored three methods: assigning from the closest joint, assigning from the closest midpoint, and assigning from the nearest joint along a hierarchy of input joints, see Figure \ref{fig:rcluster}. Our observations indicated that the latter method yielded better results. The underlying concept is that the angle formed between a vertex and a line segment passing through a joint increases as the vertex approaches the joint. Furthermore, it is important to highlight that our structure readily accommodates user input rotations.

\begin{figure*}[!htp]
    \centering
    \includegraphics[width=\textwidth]{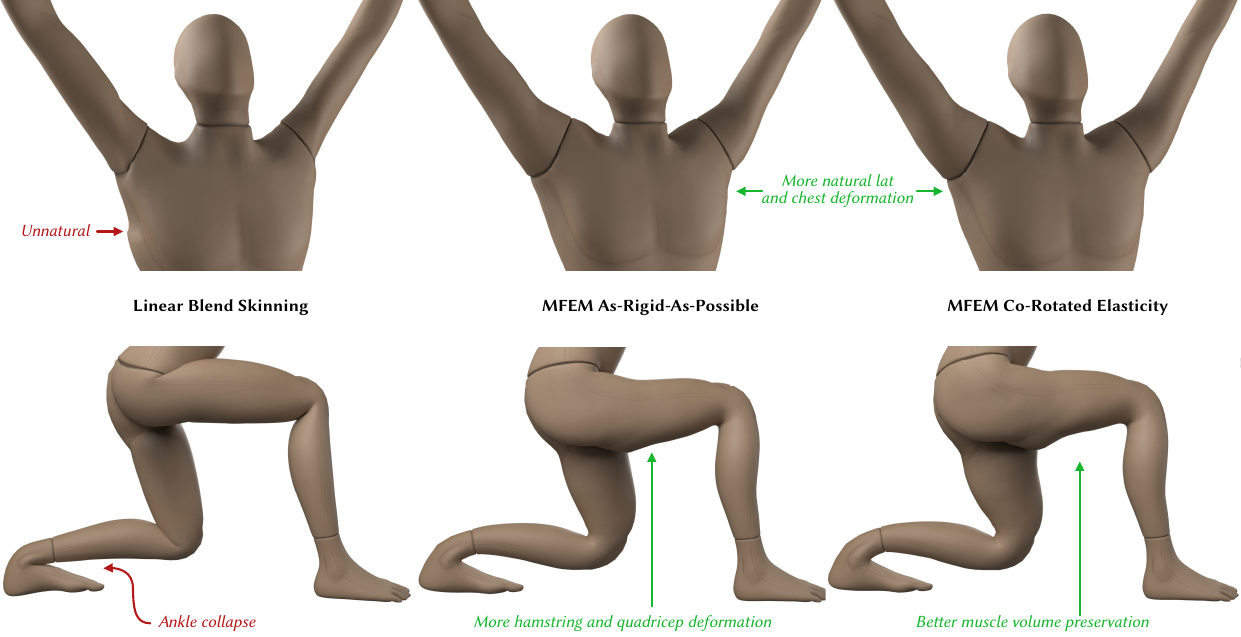}
    \caption{Using different material models with our method yields signficantly more natural deformations for the back, shoulders and lower body than standard linear blend skinning.}\label{fig:material_comparison}
\end{figure*}

\paragraph{A solver friendly to variant energies} By exposing the rotational component of the energy, which can be defined by the user, our solver can accommodate various energy formulations, as demonstrated in Figure \ref{fig:material_comparison}. We present renderings for both the ARAP energy and the co-rotational energy. Essentially, the Hessian and gradient for any material model can be incorporated into the KKT matrix. 

\paragraph{Comparing to linear blend skinning} The illustrative Figure \ref{fig:material_comparison} demonstrates that the use of MFEM preserves shape and volume better than LBS, particularly in complex skeleton hierarchies, such as when the hip/leg joint is bending. Additionally, unlike LBS, MFEM does not necessitate extensive user input, such as detailed weight painting or precise deformation definition. However, it is important to note that, in comparison to LBS, CPU-based MFEM computations are currently slower. Therefore, further experimentation and exploration of GPU implementations are necessary to address this performance gap.

\begin{figure}[H]
     \centering
     \includegraphics[width=\columnwidth]{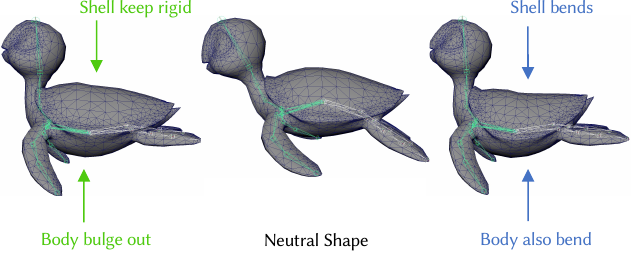}
     \caption{Using heterogeneous material on a turtle model where body material $\mu$: 1e3 and shell material $\mu$: 1e6 (left) - and homogeneous material with same material $\mu$: 1e3 of both body and shell (right).
        \label{fig:rsturtle}}
\end{figure}

\paragraph{Allowing heterogeneous material} Employing physics-based modeling allows multiple materials to be utilized by one model, which can produce physics correct poses and ease the rigging process. In Figure \ref{fig:rsturtle}, we demonstrate this by assigning different material parameters to the shell and body of a turtle. The turtle body and shell bend together with homogeneous material, while the soft body has to bulge out when the shell is extremely stiff and keeps rigid. 

\begin{figure}[H]
    \centering
    \includegraphics[width=\columnwidth]{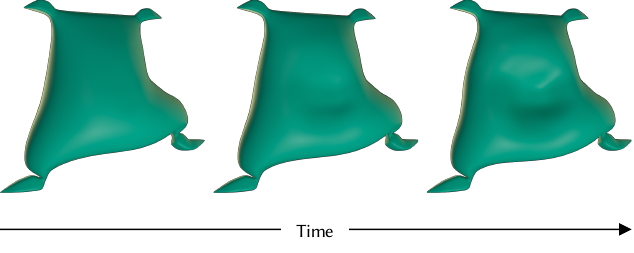}
    \caption{Smooth response to external force applied on nine vertices only of the flour sack, simulated for the co-rotational energy with $\mu = 2.4138e+06$.}\label{fig:collision_response}
\end{figure}

\paragraph{Collision response} Testing our MFEM solver against collision response. Figure \ref{fig:collision_response} shows our MFEM skinning can handle local external forces and potentially deal with local, user defined positional constraints or environmental interaction. 

\section{Conclusion Future work and Limitations}
We have presented an efficient mixed finite element skinning technique which produces physically-based deformations for character animation. Our algorithm is elegantly simple, exploiting the additional variables in mixed a discretization to enable artist control. This simplicity yields several benefits over using standard non-mixed finite element formulations for character skinning including reduction in artifacts and improved performance (via a conveniently induced linearization for practical constitutive models). We show that our method's ability to act on different material models, add additional detail such as volume effects yields results that are more natural than standard linear blend skinning approaches. Because our method does not run in a reduced space, it can react to local external  perturbations such as contacts. 

\begin{figure}[h]
    \centering
    \includegraphics[width=\columnwidth]{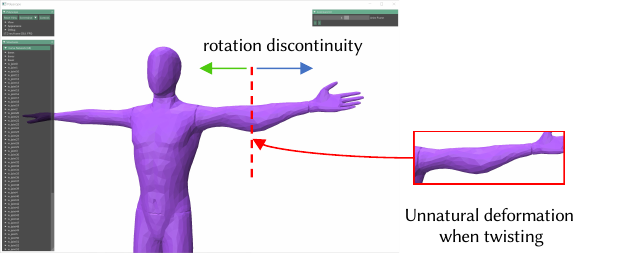}
    \caption{Large rotations discontinuities cause inflation and then eventually collapse. Similar to LBS this can be ameliorated via careful posing.}
    \label{fig:twist_artifact}
\end{figure}

The major limitations in our method stem from our fixed rotation clustering. Artifacts similar but not identical to the well-known candy wrapper effect can occur due to rotation discontinuities at cluster boundaries (Figure \ref{fig:twist_artifact}). Some additional rotation smoothing could be employed to reduce this artifacts but we leave this for future work. Further, while our method is interactive it is not as performant as LBS. Exploiting previous work on fast solvers for elasticity could help bridge this gap. Despite these issues we believe the benefits of our method, coupled with its relatively easy implementation and formulation outweigh the negatives, providing a new set of tools for generating compelling skinning based character animations without requiring tedious weight authoring or generation.

\begin{table}[h]
    \rowcolors{2}{white}{CornflowerBlue}
  \caption{Details of geometry and performance for all examples shown. Measured on a Windows PC with CPU 8-core Intel i9-11950H (2.60GHz) with 64GB RAM, Windows, python 3.10.10}
  \label{tab:freq}
  \begin{tabular}{c c c c c c}
    \hline
    Model & V & T &\text{Stiffness} & \text{Frame/Sec} \\
    \midrule
    flour sack & 2733 & 11193 & 1000.0 & 1.713\\
    human & 5922 & 21155 & 1000.0 & 1.078\\
    turtle & 4371 & 16728 & 1000.0 & 1.295\\
  \hline
\end{tabular}
\end{table}


\begin{acks}
To Abraham, to SGI, to any funding sources. Funding by the German Research Foundation (DFG) Research Training Group 2297 ”MathCoRe”, Magdeburg is gratefully acknowledged.
\end{acks}

\bibliographystyle{ACM-Reference-Format}
\bibliography{sample-base}

\end{document}